\documentclass[aps,superscriptaddress, showpacs,preprintnumbers, superscriptaddress, nofootinbibt,onecolumn, referee]{revtex4-2}
\usepackage{eurosym}
\usepackage{dcolumn}
\usepackage{bm}
\usepackage{enumerate}
\usepackage{float}
\usepackage{epstopdf}
\usepackage{amsmath}
\usepackage{bm}
\usepackage{amsfonts}
\usepackage{amssymb}
\usepackage{graphicx}
\usepackage{alphalph,mathtools}
\usepackage{etoolbox}
\usepackage{color}
\def\be{\begin{equation}}
\def\ee{\end{equation}}
\def\bea{\begin{eqnarray}}
\def\eea{\end{eqnarray}}

\begin{document}

\title{Analytical solutions of the one-dimensional Schr\"{o}dinger equation with position-dependent mass} 

\author{Tiberiu Harko}
\affiliation{Astronomical Observatory, 19 Ciresilor Street,  Cluj-Napoca 400487, Romania, }
\affiliation{Faculty of Physics, Babes-Bolyai University, 1 Kogalniceanu Street,
Cluj-Napoca 400084, Romania,}
\affiliation{School of Physics, Sun Yat-Sen University, Xingang Road, Guangzhou 510275, People's
Republic of China},
\email{tiberiu.harko@aira.astro.ro} 
\author{Man Kwong Mak}
\affiliation{Departamento de F\'{\i}sica, Facultad de Ciencias Naturales, Universidad de
Atacama, Copayapu 485, Copiap\'o, Chile}
\email{mankwongmak@gmail.com}

\begin{abstract}
The study of the Schr\"{o}dinger equation with the position-dependent
effective mass has attracted a lot of attention, due to its applications in
many fields of physics, including the properties of the semiconductors,
semiconductor heterostructures, graded alloys, quantum liquids, Helium-3
clusters, quantum wells, wires and dots etc. In the present work we obtain
several classes of solutions of the one-dimensional Schr\"{o}dinger equation
with position-dependent particle mass. As a first step the single particle
Schr\"{o}dinger equation with position-dependent mass is transformed into an
equivalent Riccati type equation. By considering some integrability cases of
the Riccati equation, seven classes of exact analytical solutions of the Schr%
\"{o}dinger equation are obtained, with the particle mass function and the
external potential satisfying some consistency conditions.
\end{abstract}

\maketitle

\tableofcontents

\section{Introduction}

The study of the position-dependent mass Schr\"{o}dinger equation (PDMSeq)
has attracted for several decades the interest of many scientists from both mathematical and physical point of view, since
PDMSeq has a considerable impact for the understanding of the behavior of quantum particles in solid
state physics, condensed matter, and related field of science \cite{r1}-\cite{r18}. Therefore, the mathematical properties of the PDMSeq and
its applications have been intensively investigated, with one of the main goals of these studies being of obtaining exact solutions of the Schr\"{o}dinger equation that includes a varying mass term.

From the mathematical point of view, the second order linear Schr\"{o}dinger equation can be reformulated
as a first order Riccati equation \cite{Ricc1,Ricc2,Ricc3} through the Cole-Hopf transformation \cite{m1,m2,m3,m4}. This representation not only allows us to obtain a deeper insight into the mathematical properties of the quantum systems, but also to obtain some exact analytical solutions of Schr\"{o}dinger equation that can be used to describe the behavior of certain quantum systems.

It is the main goal of the present paper to obtain a number of exact solutions of the one dimensional static PDMSeq by using its equivalent formulation in terms of a Riccati equation. Due to the quadratic structure of the general Riccati equation, its general
solution cannot be found easily. However, there are a number of integrability cases in which the Riccati equation can be solved exactly \cite{Ricc1,Ricc2,Ricc3,int1,int2,int3,int4}. In our present study we adopt for the PDMSeq the mathematical form introduced initially in \cite{r1}, and the corresponding ordering and kinetic operator.  Then, as a first step in our investigation we reformulate the PDMSeq as a first order nonlinear Riccati type equation, the PDMSReq. Then, by using some integrability conditions for the PDMSReq, we obtain seven classes of solutions of the Schr\"{o}dinger equation with position-dependent mass, with the mass function of the quantum particle and the external potential satisfying some integrability conditions.

The present paper is organized as follows. The Schr\"{o}dinger equation with position-dependent mass with the von Roos kinetic term is presented in Section~\ref{sect1}, where the equivalent Ricacti equation is also obtained. Several solutions of the PDMSeq are obtained in Section~\ref{sect2}, by imposing some constraints on the coefficients of the Riccati equation.  Solutions of the PDMSeq depending on an arbitrary function and with the mass function and the potential satisfying some differential and integral conditions are obtained in Sections~\ref{sect3} and \ref{sect4}, respectively. We discuss and conclude our results in Section~\ref{sect5}.

\section{The Schr\"{o}dinger equation with position-dependent mass}\label{sect1}

In standard nonrelativistic quantum mechanics the kinetic energy operator takes
the simple form $\hat{T}_{s}=\hat{p}^{2}/2m$,
 where $m$ is the (constant) particle mass, and $\hat{p}
$ is the momentum operator. However, this kinetic operator
is ill-defined for a particle with position-dependent mass $m=m(x)$.   Thus, for the case of PDM one has to determine how to order the mass
relative to the momentum operators in order to generalize the standard
kinetic energy operator $\hat{T}_{s}$, and to construct the PDMSeq. In the following we will discuss and adopt the
generalized kinetic operator introduced initially by von Roos \cite{r1}, and which solves
the PDM ordering problem in a simple way.

Let's consider the Hermitian kinetic operator given by \cite{r1}
\begin{eqnarray}\label{T1}
\hat{T}&=&\frac{1}{8}\Bigg\{\left[ m^{-1}\left( \vec{r}\right) \hat{p}^{2}+\hat{p}%
^{2}m^{-1}\left( \vec{r}\right) \right] +m^{\alpha }\left( \vec{r}\right)
\hat{p}m^{\beta }\left( \vec{r}\right) \hat{p}m^{\gamma }\left( \vec{r}%
\right) + \nonumber\\
&&m^{\gamma }\left( \vec{r}\right) \hat{p}m^{\beta }\left( \vec{r}%
\right) \hat{p}m^{\alpha }\left( \vec{r}\right)\Bigg\},
\end{eqnarray}
where $\alpha $, $\beta $ and $\gamma $ are ambiguous
parameters satisfying the constraint
\begin{equation}
\alpha +\beta +\gamma =-1,
\end{equation}
as proposed in \cite{r1},  The first term in the above equation is added here in order to include in the general expression the usual
symmetrized or Weyl ordered operator \cite{9}. In
one dimension, we substitute the momentum operator $\hat{p}$ according to $%
\hat{p}=-i\hbar \frac{d}{dx}$, where $\hbar$ is Planck's constant,  to the right hand side of Eq.~(\ref{T1}) yielding
\begin{equation}
\hat{T}=\frac{1}{2m(x)}\frac{d^{2}}{dx^{2}}+\frac{i\hbar }{2m(x)}\frac{d\ln m(x)}{dx}%
\frac{d}{dx}+U_{k}\left( x\right) ,  \label{T2}
\end{equation}
where we have introduced an arbitrary, mass dependent function $U_{k}\left( x\right) $, playing the role of an effective potential, and
defined as
\begin{equation}\label{T3}
U_{k}\left( x\right) =\frac{\hbar ^{2}}{4m^{3}(x)}\left[ \left( 1-\alpha
-\gamma \right) \frac{m(x)}{2}\frac{d^{2}m(x)}{dx^{2}}+\left( \alpha \gamma
+\alpha +\gamma -1\right) \left( \frac{dm(x)}{dx}\right) ^{2}\right] ,
\end{equation}

The effective potential $U_{k}\left( x\right) $ vanishes subject to the
conditions
\begin{equation}
1-\alpha -\gamma =\alpha \gamma +\alpha +\gamma -1=0,  \label{T4}
\end{equation}
which imply that $\alpha =0$ and $\gamma =1$, or $\alpha =1$ and $\gamma =0$, respectively.
Note that the kinetic operator $\hat{T}$ is free of the uncertainties coming
from the commutation rules of quantum mechanics $\hat{x}\hat{p}-\hat{p}\hat{x%
}=i\hbar$.

For an arbitrary external potential $V\left( x\right) $, the
non-ambiguous PDM Schr\"{o}dinger equation takes the form
\begin{equation}
\left[ \frac{1}{2m(x)}\hat{p}^{2}+\frac{i\hbar }{2m(x)}\frac{d\ln m(x)}{dx}\hat{p}%
+V\left( x\right) \right] \psi \left( x\right) =E\psi \left( x\right) .
\label{T5}
\end{equation}

By taking into account the explicit expression of the momentum operator Eq.~(\ref{T5}) can be written as a
second order differential equation,
\begin{equation}  \label{T6}
\frac{d^{2}\psi (x)}{dx^{2}}-\frac{d\ln m(x)}{dx}\frac{d\psi (x)}{dx}+\frac{%
2m(x)}{\hbar ^{2}}\left[ E-V\left( x\right) \right] \psi (x)=0.
\end{equation}

In order to solve the PDM Schr\"{o}dinger Eq.~(\ref{T6}), we introduce the
auxiliary function $u(x)$, defined according to
\begin{equation}
\psi \left( x\right) =\psi _{0}e^{\int^{x}u(\phi )d\phi },
\end{equation}%
where $\psi _{0}$ is an arbitrary constant. With the help of this transformation, Eq.~(%
\ref{T6}) reduces to the standard Riccati type differential equation
\begin{equation}
\frac{du}{dx}=a\left( x\right) +b\left( x\right) u+c\left( x\right) u^{2},
\label{T7}
\end{equation}%
where we have denoted
\begin{equation}
a\left( x\right) =\frac{2m(x)}{\hbar ^{2}}\left[ V\left( x\right) -E\right],
\end{equation}
\begin{equation}
 b\left( x\right) =\frac{d\ln m}{dx},
 \end{equation}
 \begin{equation}
  c\left( x\right) =-1.
\end{equation}

\section{Integrability cases for the PDM Schr\"{o}dinger equation}\label{sect2}

In the following we consider some integrability cases of the PDM Schr\"{o}%
dinger-Riccati Eq.~(\ref{T7}), which allow us to obtain exact
analytical solutions of the PDMSeq.

\subsection{Case 1: $a(x)+\beta b(x)+\beta ^{2}c(x)=0,%
\beta =\mathrm{constant}$}

We assume first that the coefficients $a(x)$, $b(x)$ and $c(x)$ of Eq.~(\ref%
{T7}) satisfy the condition
\begin{equation}
a(x)+\beta b(x)+\beta ^{2}c(x)=0,  \label{cond2}
\end{equation}
where $\beta $ is an arbitrary constant. Explicitly, Eq.~(\ref{cond2}) takes
the form
\begin{equation}  \label{cond2a}
\frac{2m(x)}{\hbar ^{2}}\left[ V\left( x\right) -E\right] +\beta \frac{d\ln m(x)%
}{dx}=\beta ^{2}.
\end{equation}%

By means of the substitution
\begin{equation}
u(x)=\beta +v(x),
\end{equation}
it follows that the
function $v(x)$ satisfies the Bernoulli type equation
\begin{equation}
\frac{dv}{dx}=\left[ b(x)-2\beta \right] v(x)-v^{2}(x),
\end{equation}%
with the general solution given by
\begin{equation}
v(x)=\frac{e^{-\int^{x}\left[ 2\beta -b(\phi )\right] d\phi }}{%
C_{1}+\int^{x}e^{-\int^{\psi }\left[ 2\beta -b(\phi )\right] d\phi }d\psi }=%
\frac{d}{dx}\ln \left| C_{1}+\int^{x}e^{-\int^{\psi }\left[ 2\beta -b(\phi )%
\right] d\phi }d\psi \right| ,
\end{equation}%
where $C_{1}$ is an arbitrary constant of integration. Therefore we have
obtained the following

\textbf{Theorem 1}. \textit{If the position-dependent mass $m(x)$ of a
quantum particle and the external potential $V(x)$ satisfy the condition
given by Eq.~(\ref{cond2a}), then the general solution of the PDM Schr\"{o}%
dinger equation is given by
\begin{equation}
\psi (x)=\psi _{0}e^{\beta x}\left[ C_{1}+\int^{x}m(\phi )e^{-2\beta \phi
}d\phi \right] .
\end{equation}
} The potential $V(x)$ can be obtained from the known mass distribution $m(x)
$ as
\begin{equation}
V(x)=E+\frac{\beta \hbar ^{2}}{2m(x)}\left[ \beta -\frac{d\ln m(x)}{dx}\right] ,
\end{equation}%
while the mass distribution can be expressed as a function of the potential
in the form
\begin{equation}
m(x)=\frac{e^{\beta x}}{m_{2}+\left( 2/\beta \hbar ^{2}\right) \int^{x}{%
e^{\beta \phi }\left[ V(\phi )-E\right] d\phi }},
\end{equation}%
where $m_{2}$ is an arbitrary constant of integration.

By considering a mass
function of the form
\begin{equation}
m(x)=m_{0}\mathrm{sech}^{2}(\omega x),
\end{equation}
with $m_{0}$ and $\omega \neq 1$ constants, we obtain the potential as
\begin{equation}
V(x)=E+\frac{\beta \hbar ^{2}\cosh ^{2}(\omega x)\left[ \beta +2\omega \tanh
(\omega x)\right] }{2m_{0}}.
\end{equation}

The wave function that solves the Schr\"{o}dinger equation for this mass
distribution and potential is given by
\begin{eqnarray}
\hspace{-0.4cm}\psi (x)&=&\frac{\psi _{0}}{\omega ^{2}}e^{-\beta x}\Bigg\{ %
\omega ^{2}C_{1}e^{2\beta x}+m_{0}\omega \tanh (\omega x)+  \nonumber \\
\hspace{-0.4cm}&&\left( -1\right) ^{1+\frac{\beta }{\omega }}\beta
m_{0}e^{2\beta x}\left[ B_{-e^{2\omega x}}\left( 1-\frac{\beta }{\omega }%
,0\right) +B_{-e^{2\omega x}}\left( -\frac{\beta }{\omega },0\right) \right] %
\Bigg\} ,
\end{eqnarray}
where $\psi _0$ is an arbitrary integration constant, and $B_{z}\left(
a,b\right) $ denotes the incomplete beta functions, defined as \cite{AS}
\begin{equation}
B_{z}(a,b)=\int_{0}^{z}t^{a-1}\left( 1-t\right) ^{b-1}dt.
\end{equation}

\subsection{Case 2: $a(x)=a_{0}^{2}e^{2\int^{x}{b(\phi )d%
\phi }}$}

If the coefficients $a(x)$ and $b(x)$ of the Riccati Eq.~(\ref{T7}), with $%
c(x)=-1$, satisfy the condition
\begin{equation}
b(x)=\frac{1}{2}\frac{d\ln a(x)}{dx},
\end{equation}%
or, equivalently,
\begin{equation}
a(x)=a_{0}^{2}e^{2\int^{x}{b(\phi )d\phi }},
\end{equation}
where $a_{0} $ is an arbitrary constant of integration, the Riccati Eq.~(\ref{T7})
takes the form
\begin{equation}
\frac{du(x)}{dx}=a(x)+\frac{1}{2}\frac{d\ln a(x)}{dx}u(x)-u^{2}(x).
\label{ricc3}
\end{equation}%
With the help of the transformation $u(x)=\sqrt{a(x)}f(x)$, Eq.~(\ref{ricc3}%
) becomes
\begin{equation}
\frac{df}{dx}=\sqrt{a(x)}\left[ 1-f^{2}(x)\right] ,
\end{equation}%
with the general solution given by
\begin{equation}
f(x)=\tanh \left[ \int^{x}{\sqrt{a\left( \phi \right) }d\phi }+f_{0}\right] ,
\end{equation}%
where $f_{0}$ is an arbitrary constant of integration.

Therefore we have obtained the following

\textbf{Theorem 2}. \textit{If the position-dependent mass $m(x)$ of a
quantum particle and the external potential $V(x)$ satisfy the condition,
\begin{equation}
a_{0}^{2}m(x)=\frac{2}{\hbar ^{2}}\left[ V(x)-E\right] ,
\end{equation}%
then the general solution of the Schr\"{o}dinger equation is given by
\begin{eqnarray}
\psi (x)&=&\psi _{0}\cosh \left[ \frac{\sqrt{2}}{\hbar }\int^{x}{\sqrt{%
m(\phi )\left[ V\left( \phi \right) -E\right] }d\phi }+f_{0}\right] =
\nonumber \\
&& a_{0}\int^{x}{m(\phi )d\phi }+f_{0} .
\end{eqnarray}
}

\subsection{Case 3: $b^{2}(x)-2\frac{db(x)}{dx}+4a(x)=\Delta =\mathrm{%
constant}$}

We assume now that the coefficients $a(x)$, $b(x)$ and $c(x)$ of the PDMSReq Eq.~(%
\ref{T7}) satisfy the condition
\begin{equation}
b^{2}(x)-2\frac{db}{dx}+4a(x)=\Delta ,  \label{qq}
\end{equation}
where $\Delta $ is an arbitrary constant. Equivalently, Eq.~(\ref{qq}) takes
the explicit form%
\begin{equation}  \label{cond3}
V(x)=E+\frac{\hbar ^{2}}{8m(x)}\left[ \Delta +2\frac{d^{2}\ln m(x)}{dx^{2}}%
-\left( \frac{d\ln m(x)}{dx}\right) ^{2}\right] .
\end{equation}
%\begin{equation}
%\left( \frac{d\ln m(x)}{dx}\right) ^{2}-2\frac{d^{2}\ln m(x)}{dx^{2}}+\frac{8m(x)}{%
%\hbar ^{2}}\left[ V\left( x\right) -E\right] =\Delta .  \label{cond3}
%\end{equation}%
As one can check by direct calculations, the general solution of the Riccati
Eq. (\ref{T7}) with coefficients satisfying condition (\ref{qq}) is given by
\begin{equation}
u_{\pm }=\frac{1}{2}b(x)\pm \frac{\sqrt{\Delta }}{2}=\frac{1}{2}\frac{d\ln
m(x)}{dx}\pm \frac{\sqrt{\Delta }}{2}.
\end{equation}

Therefore we have obtained the following

\textbf{Theorem 3}. \textit{If the position-dependent mass $m(x)$ and the
potential $V(x)$ satisfy the condition given by Eq.~(\ref{cond3}), then the
general solution of the Schr\"{o}dinger equation takes the form
\begin{equation}
\psi _{\pm }(x)=\psi _{0\pm }e^{\pm \sqrt{\Delta }x/2}\sqrt{m(x)},
\end{equation}
where $\psi _{0\pm }$ are the arbitrary constants of integration. }

As a function of the potential the mass distribution $m(x)$ satisfies the
second order differential equation
\begin{equation}
\frac{2}{m(x)}\frac{d^{2}m(x)}{dx^{2}}-3\left( \frac{d\ln m(x)}{dx}\right)
^{2}-\frac{8m(x)}{\hbar ^{2}}\left[ V(x)-E\right] +\Delta =0.  \label{deq1}
\end{equation}%
The transformation $m(x)M^{2}(x)=1$ reduces Eq.~(\ref{deq1}) to the form
\begin{equation}
\frac{d^{2}M(x)}{dx^{2}}-\frac{\Delta }{4}M(x)+\frac{2}{\hbar ^{2}}\left[
V(x)-E\right] \frac{1}{M(x)}=0.
\end{equation}

\section{Solutions of the PDM Schr\"{o}dinger equation depending on
arbitrary functions}\label{sect3}

In the present Section we will consider solutions of the Schr\"{o}%
dinger-Riccati Eq.~(\ref{T7}) that \textit{depend on an arbitrary function $%
f\left( x\right) \in C^{\infty }(I)$ defined on a real interval $I\subseteq
\Re $.} In the following we explicitly take into account that $c(x)\equiv -1$%
. The starting point of our analysis is the requirement that Eq.~(\ref{T7})
admits particular solutions of the form
\begin{equation}
u_{\pm }^{(p)}(x)=\frac{b(x)}{2}\pm \frac{\sqrt{f(x)}}{2}.
\end{equation}
For Eq.~(\ref{T7}) to admit such solutions, its coefficients $a(x)$ and $b(x)
$ must satisfy the consistency condition \cite{int1,int2}
\begin{equation}
a(x)=\frac{d}{dx}\left[ \frac{b(x)}{2}\pm \frac{\sqrt{f(x)}}{2}\right] -%
\frac{b^{2}(x)-f(x)}{4}.  \label{cond5}
\end{equation}

By introducing a new function $v(x)$, so that
\begin{equation}
u(x)=u_{\pm }^{(p)}(x)+v(x),
\end{equation}
it follows that the function $v(x)$ satisfies the Bernoulli type equation
\begin{equation}
\frac{dv}{dx}\pm \sqrt{f(x)}v(x)+v^{2}(x)=0,
\end{equation}%
with the general solutions given by
\begin{equation}
v_{\pm }(x)=\frac{e^{\mp \int^{x}{\sqrt{f(\phi )}d\phi }}}{%
v_{0}+\int^{x}e^{\mp \int^{\psi }{\sqrt{f(\phi )}d\phi }}{d\psi }}=\frac{d}{%
dx}\ln \left| v_{0}+\int^{x}e^{\mp \int^{\psi }{\sqrt{f(\phi )}d\phi }}{%
d\psi }\right| ,
\end{equation}
where $v_{0}$ is an arbitrary constant of integration. Therefore we have
obtained the following

\textbf{Theorem 4}. \textit{If the position-dependent mass $m(x)$ and the
external potential $V(x)$ satisfy the condition
\begin{equation}
\frac{4m(x)}{\hbar ^{2}}\left[ V\left( x\right) -E\right] =\frac{d^{2}\ln m(x)}{%
dx^{2}}-\frac{1}{2}\left[ \frac{d\ln m(x)}{dx}\right] ^{2}+\frac{f(x)}{2}\pm
\frac{d}{dx}\sqrt{f(x)},
\end{equation}%
where $f(x)$ is an arbitrary function of the independent variable $x$, then
the general solution of the Schr\"{o}dinger equation is given by
\begin{equation}
\psi _{\pm }(x)=\psi _{0}\sqrt{m(x)}e^{\pm \frac{1}{2}\int^{x}\sqrt{f(\phi )}%
d\phi }\left\{ v_{0}+\int^{x}e^{\mp \int^{\psi }{\sqrt{f(\phi )}d\phi }}{%
d\psi }\right\} .
\end{equation}
} Some particular integrability cases can be easily obtained from \textbf{%
Theorem 4}.

\subsection{Particular case 4a: $f(x)=0$}

If the arbitrary function $f(x)$ vanishes, then it immediately follows that
if the mass and the potential of the PDM Schr\"{o}dinger equation satisfy
the constraint
\begin{equation}
\frac{4m(x)}{\hbar ^{2}}\left[ V\left( x\right) -E\right] =\frac{d^{2}\ln
m(x)}{dx^{2}}-\frac{1}{2}\left[ \frac{d\ln m(x)}{dx}\right] ^{2},
\end{equation}%
then the general solution of the Schr\"{o}dinger equation is given by
\begin{equation}
\psi (x)=\psi _{0}\sqrt{m(x)}\left( v_{0}+x\right) .
\end{equation}

If $f(x)$ is an arbitrary constant, then we recover \textbf{Theorem 3}.

\subsection{Particular case 4b: $f(x)=b^{2}(x)$}

With the particular choice $f(x)=b^{2}(x)$, by taking the plus sign in Eq. (%
\ref{cond5}), the integrability condition of the PDM Schr\"{o}dinger
equation becomes
\begin{equation}
\frac{2m(x)}{\hbar ^{2}}\left[ V\left( x\right) -E\right] =\frac{d^{2}\ln m
(x)}{dx^{2}},
\end{equation}%
with the general solution of the Schr\"{o}dinger equation given by
\begin{equation}
\psi (x)=\psi _{0}m(x)\left[ v_{0}+\int^{x}\frac{d\phi }{m(\phi )}\right] .
\end{equation}

\section{Further integrability cases for the PDM Schr\"{o}dinger-Riccati
equation}\label{sect4}

Using the algorithm and the \textbf{Theorems} obtained in \cite%
{int4}, in the following we will obtain \textbf{Theorems} 5 to 7, giving the solutions of the PDMSeq for a specific set of constraints satisfied by the mass function and the external potential, also depending on an arbitrary function.   In order to make the present paper readable, and to avoid the
repetitive calculations, we shall not present the full demonstrations for
obtaining the following {\bf Theorems}, and we refer the reader to \cite{int4} for the mathematical details.

The basic idea of the mathematical approach used in the present Section is as follows. The Riccati Eq.~(\ref{T7}) is a
second order algebraic equation in $u\left( x\right) $. Now we assume that its
particular solutions $u_{\pm }^{p}(x)$ are given by
\begin{equation}\label{mn1}
u_{\pm }^{p}(x)=\frac{-b\left( x\right) \pm \sqrt{%
b^{2}\left( x\right) -4a(x)c(x)+4c(x)\frac{du^{p}(x)}{dx}}}{2c(x)}.
\end{equation}

To obtain the general solution of the Riccati Eq.~(\ref{T7}%
) by using Eq.~(\ref{mn1}), we introduce the solution generating
function $f\left( x\right) $, which satisfies the first order differential
equation in $u^p(x)$, given by
\begin{equation}
b^{2}\left( x\right) +4c(x)\frac{du^{p}(x)}{dx}%
=f\left( x\right).
\end{equation}

If the above equation can be integrated we obtain the explicit form of a
particular solution of the Riccati Eq.~(\ref{T7}). Hence,  since its particular
solution is known, the general solution of the Riccati Eq.~(\ref{T7}) can be obtained through quadratures.

\subsection{Case 4: $a\left( x\right) =\frac{1}{2}\frac{d}{dx}\left[ \frac{%
d\ln m(x)}{dx}\mp \sqrt{f\left( x\right) +\left( \frac{d\ln m(x)}{dx}%
\right) ^{2}}\right] +\frac{f\left( x\right) }{4}$}

We assume now that $a\left( x\right) $ satisfies the differential condition%
\begin{equation}  \label{d1}
a\left( x\right) =\frac{1}{2}\frac{d}{dx}\left[ \frac{d\ln m(x)}{dx}\mp
\sqrt{f\left( x\right) +\left( \frac{d\ln m(x)}{dx}\right) ^{2}}\right] +%
\frac{f\left( x\right) }{4},
\end{equation}%
where $f(x)$ is an arbitrary function. By substituting Eq.~(\ref{d1}) into
the Riccati Eq.~(\ref{T7}), the latter can be expressed as
\begin{eqnarray}  \label{TH2}
\frac{du_{\mp }(x)}{dx}&=&\frac{1}{2}\frac{d}{dx}\left[ \frac{d\ln m(x)}{dx}%
\mp \sqrt{f\left( x\right) +\left( \frac{d\ln m(x)}{dx}\right) ^{2}}\right] +%
\frac{f\left( x\right) }{4}+  \nonumber \\
&&\frac{d\ln m(x)}{dx}u_{\mp }(x)-u_{\mp }^{2}(x).
\end{eqnarray}

Therefore we obtain the following:

\textbf{Theorem 5}. \textit{If the position-dependent mass $m(x)$ and the
potential $V(x)$ satisfy the condition
\begin{equation}
\frac{2m(x)}{\hbar ^{2}}\left[ V\left( x\right) -E\right] =\frac{1}{2}\frac{d%
}{dx}\left[ \frac{d\ln m(x)}{dx}\mp \sqrt{f\left( x\right) +\left( \frac{d\ln
m(x)}{dx}\right) ^{2}}\right] +\frac{f\left( x\right) }{4},
\end{equation}%
where $f(x)$ is an arbitrary function of the independent variable $x$, then
the general solutions of the Schr\"{o}dinger equation are given by
\begin{eqnarray}
\psi _{\mp }(x)&=&\psi _{0}\sqrt{m(x)}\left[ C_{\mp }+\int^{x}e^{-\int^{\psi
}\sqrt{f\left( \phi \right) +\left( \frac{d\ln m(\phi)}{d\phi }\right) ^{2}}%
d\phi }d\psi \right] \times  \nonumber \\
&&\exp \left[ \mp \int^{x}\sqrt{f\left( \phi \right) +\left( \frac{d\ln m
(\phi)}{d\phi }\right) ^{2}}d\phi \right] ,  \label{s2}
\end{eqnarray}
where $C_{\mp }$ are arbitrary constants of integration. }

\subsection{Case 5: $a\left( x\right) =\frac{1}{4}\left\{ f(x)\left[ 2\frac{%
d\ln m(x)}{dx}+f(x)\right] -\frac{df(x)}{dx}\right\} $}

Assume now that the arbitrary function $a(x)$ satisfies the condition%
\begin{equation}
a\left( x\right) =\frac{1}{4}\left\{ f(x)\left[ 2\frac{d\ln m(x)}{dx}+f(x)%
\right] -\frac{df(x)}{dx}\right\} ,  \label{d4}
\end{equation}%
where $f(x)$ is an arbitrary function. By substituting Eq.~(\ref{d4}) into
the Riccati Eq.~(\ref{T7}), the latter can be expressed as
\begin{equation}
\frac{du(x)}{dx}=\frac{1}{4}\left\{ f(x)\left[ 2\frac{d\ln m(x)}{dx}+f(x)%
\right] -\frac{df\left( x\right) }{dx}\right\} +\frac{d\ln m(x)}{dx}%
u(x)-u^{2}(x).  \label{TH6}
\end{equation}

Therefore we obtain the following

\textbf{Theorem 6}. \textit{If the position-dependent mass $m(x)$ of a
quantum particle and the external potential $V(x)$ satisfy the condition
\begin{equation}
\frac{2m(x)}{\hbar ^{2}}\left[ V\left( x\right) -E\right] =\frac{1}{4}%
\left\{ f(x)\left[ 2\frac{d\ln m(x)}{dx}+f(x)\right] -\frac{df(x)}{dx}%
\right\} ,
\end{equation}
then the general solutions of the Schr\"{o}dinger equation are given by
\begin{eqnarray}
\psi (x)&=&\psi _{0}\sqrt{m(x)}\left\{ C+\int^{x}e^{\int^{\omega }\sqrt{%
\left[ \frac{d\ln m(\psi )}{d\psi }\right] ^{2}+f\left( \psi \right) \left[ 2%
\frac{d\ln m(\psi )}{d\psi }+f\left( \psi \right) \right] }d\psi }d\omega
\right\}\times  \nonumber \\
&&\exp \left\{ -\frac{1}{2}\int^{x}\sqrt{\left[ \frac{d\ln m(\phi )}{d\phi }%
\right] ^{2}+f(\phi )\left[ 2\frac{d\ln m(\phi )}{d\phi }+f(\phi )\right] }%
d\phi \right\} .
\end{eqnarray}
}

\subsection{Case 6: $b\left( x\right) =\frac{d\ln f(x)}{dx}-\frac{f\left(
x\right) }{2}+\frac{4}{\hbar ^{2}}\frac{\left[ V\left( x\right) -E\right]
e^{-\frac{1}{2}\int^{x}f\left( \phi \right) d\phi }}{%
C_{5}-\frac{4}{\hbar ^{2}}\int^{x}\left[ V\left( \psi %
\right) -E\right] f\left( \psi \right) e^{-\frac{1}{2}\int^{%
\psi }f\left( \phi \right) d\phi }d\psi }$}

We assume that the coefficient $b\left( x\right) $ of the Riccati Eq.~(\ref%
{T7}) satisfies the differential condition
\begin{equation}
b\left( x\right) =\frac{d\ln f(x)}{dx}-\frac{f\left( x\right) }{2}+\frac{4}{%
\hbar ^{2}}\frac{\left[ V\left( x\right) -E\right] e^{-\frac{1}{2}%
\int^{x}f\left( \phi \right) d\phi }}{C_{5}-\frac{4}{\hbar ^{2}}%
\int^{x}f\left( \psi \right) \left[ V\left( \psi \right) -E\right] e^{-\frac{%
1}{2}\int^{\psi }f\left( \phi \right) d\phi }d\psi },  \label{b8}
\end{equation}%
where $C_{5}$ is an arbitrary constant, and $f(x)$ an arbitrary function. By
substituting Eq.~(\ref{b8}) into the Riccati Eq.~(\ref{T7}), the latter can
be expressed as
\begin{equation}\label{53}
\begin{split}
 \frac{du(x)}{dx}&=\frac{2}{\hbar ^{2}}\frac{f\left(
x\right) \left[ V\left( x\right) -E\right] e^{-\frac{1}{2}\int^{x}f\left(
\phi \right) d\phi }}{C_{5}-\frac{4}{\hbar ^{2}}\int^{x}f\left( \psi \right) %
\left[ V\left( \psi \right) -E\right] e^{-\frac{1}{2}\int^{\psi }f\left(
\phi \right) d\phi }d\psi }+  \\
 &\left\{ \frac{d\ln f(x)}{dx}-\frac{f\left( x\right) }{2}+%
\frac{4}{\hbar ^{2}}\frac{\left[ V\left( x\right) -E\right] e^{-\frac{1}{2}%
\int^{x}f\left( \phi \right) d\phi }}{C_{5}-\frac{4}{\hbar ^{2}}%
\int^{x}f\left( \psi \right) \left[ V\left( \psi \right) -E\right] e^{-\frac{%
1}{2}\int^{\psi }f\left( \phi \right) d\phi }d\psi }\right\}\times \\
& u(x)- u^{2}(x).
\end{split}
\end{equation}

Therefore we obtain the following

\textbf{Theorem 7}. \textit{If the coefficient $b(x)$ of the Schr\"{o}dinger-Riccati Eq.~(%
\ref{T7}) satisfies the differential condition~(\ref{b8}), then the general
solution of the Riccati Eq.~(\ref{53}) is given by
\begin{equation}\label{TH8a}
\begin{split}
 u(x)&=    \\
 &\frac{e^{\int^{x}\left\{ \left\{ \frac{d\ln f(\omega)}{d\omega }%
-\frac{f\left( \omega \right) }{2}+\frac{4}{\hbar ^{2}}\frac{\left[ V\left(
\omega \right) -E\right] e^{-\frac{1}{2}\int^{\omega }f\left( \phi
\right) d\phi }}{C_{5}-\frac{4}{\hbar ^{2}}\int^{\omega }f\left( \psi
\right) \left[ V\left( \psi \right) -E\right] e^{-\frac{1}{2}\int^{\psi
}f\left( \phi \right) d\phi }d\psi }\right\} +f\left( \omega \right)
\right\} d\omega }}{C_{6}+\int^{x}e^{\int^{\eta }\left\{ \left\{ \frac{d\ln
f(\omega)}{d\omega }-\frac{f\left( \omega \right) }{2}+\frac{4}{\hbar ^{2}}%
\frac{\left[ V\left( \omega \right) -E\right] e^{-\frac{1}{2}\int^{\omega
}f\left( \phi \right) d\phi }}{C_{5}-\frac{4}{\hbar ^{2}}\int^{\omega
}f\left( \psi \right) \left[ V\left( \psi \right) -E\right] e^{-\frac{1}{%
2}\int^{\psi }f\left( \phi \right) d\phi }d\psi }\right\} +f\left(
\omega \right) \right\} d\omega }d\eta }   \\
 &-\frac{f\left( x\right) }{2},
 \end{split}
\end{equation}%
where $C_{6}$ is an arbitrary constant of integration. }

The position-dependent mass $m\left( x\right) $ must satisfy the condition
\begin{equation}
m\left( x\right) =\frac{f\left( x\right) e^{-\frac{1}{2}\int^{x}f\left( \phi
\right) d\phi }}{C_{5}-\frac{4}{\hbar ^{2}}\int^{x}f\left( \psi \right) %
\left[ V\left( \psi \right) -E\right] e^{-\frac{1}{2}\int^{\psi }f\left(
\phi \right) d\phi }d\psi }.
\end{equation}

Using the general solutions of the Riccati Eq.~(\ref{T7}) given by Eqs. (\ref%
{TH8a}), the general solutions of the Schr\"{o}dinger Eq.~(\ref{T6}) can be
obtained from $\psi \left( x\right) =\psi _{0}e^{\int^{x}u\left( \phi
\right) d\phi }$, respectively.

\section{Conclusions}\label{sect5}

In the present paper we have obtained several classes of exact solutions of the one-dimensional PDMSeq with the von Roos kinetic operator \cite{r1}, by using its equivalent mathematical representation in terms of a Riccati equation. By imposing some functional relations between the coefficients of the Riccati equation one can obtain the exact solution of the PDMSeq. The obtained solutions can be classified into two classes. In the first class the integrability condition depends only on the mass function, the external potential, and the energy of the system. In the second class of solutions the integrability condition, as well as the wave function obtained as the solution of the PDMSeq, depends on an arbitrary function $f(x)$. The integrability conditions for the second class of solutions take some very complicated differential/integral forms.

Generally, from the integrability conditions one can express the external potential as $V=V\left(E,m(x), f(x)\right)$, that is, as a function of the energy, mass, and an arbitrary function. Hence all the obtained solutions belong to the class of the systems with energy dependent potentials, a class of models that have found many applications in physics (for a discussion of the role of energy dependent potentials in quantum mechanics see \cite{31} and references therein). It is important note that if the external potential and the wave function are  energy dependent, in the Hilbert space the norm (scalar product) of $\psi$ is defined according to \cite{31}
\begin{equation}
N=\int_{-\infty }^{+\infty }{\psi ^{\ast }(x,y)\left[ 1-\frac{\partial
V\left( x,y,E\right) }{\partial E}\right]\psi (x,y) dxdy}>0.
\end{equation}
Moreover,  in microscopic systems in the presence of energy-dependent potential the standard quantum mechanically
completeness relation $\sum_{n}\psi _{n}\left( x^{\prime },y^{\prime
}\right) \psi _{n}^{\ast }\left( x,y\right) =\delta \left( x-x^{\prime
}\right) \delta \left( y-y^{\prime }\right) $ is not valid generally.

On the other hand, different external potentials can generate a large variety of position-dependent mass functions in quantum mechanical structures. Hopefully, the obtained results may find some applications in the study of various condensed matter systems, in the presence of physically relevant external potentials. The physical applications of the obtained solutions will be presented in a future work.

\section*{Acknowledgments} 

We would like to thank the anonymous reviewer for comments and suggestions that helped us to improve our manuscript.

\end{document}